\pdfoutput=1
\documentclass{article}

\usepackage{microtype}
\usepackage{graphicx}
\usepackage{booktabs} 

\usepackage{hyperref}

\usepackage{amsmath}
\usepackage{amssymb}
\usepackage{amsfonts}
\usepackage{bbm}
\usepackage{bm}
\usepackage{xcolor}
\usepackage{todonotes}
\usepackage{caption}
\usepackage{subcaption}
\usepackage{makecell}
\usepackage{tablefootnote}

\DeclareCaptionStyle{ruled}{labelfont=normalfont,labelsep=colon,strut=off}
\usepackage[accepted]{mlsys2023}

\mlsystitlerunning{MUSTACHE: Multi-Step-Ahead Predictions for Cache Eviction}

\begin{document}

\twocolumn[
\mlsystitle{MUSTACHE: Multi-Step-Ahead Predictions for Cache Eviction}



\mlsyssetsymbol{equal}{*}

\begin{mlsysauthorlist}
\mlsysauthor{Gabriele Tolomei}{sapienza}
\mlsysauthor{Lorenzo Takanen}{sapienza}
\mlsysauthor{Fabio Pinelli}{imt}
\end{mlsysauthorlist}

\mlsysaffiliation{sapienza}{Department of Computer Science, Sapienza University of Rome, Italy}
\mlsysaffiliation{imt}{IMT School for Advanced Studies, Lucca, Italy}

\mlsyscorrespondingauthor{Gabriele Tolomei}{tolomei@di.uniroma1.it}

\mlsyskeywords{Cache replacement algorithm, learning to evict cache entries}

\vskip 0.3in

\begin{abstract}

In this work, we propose MUSTACHE, a new page cache replacement algorithm whose logic is 
{\em learned} from observed memory access requests rather than fixed like existing policies.
We formulate the page request prediction problem as a categorical time series forecasting task. 
Then, our method queries the learned page request forecaster to obtain the next $k$ predicted page memory references to better approximate the optimal Bélády's replacement algorithm.
We implement several forecasting techniques using advanced deep learning architectures and integrate the best-performing one into an existing open-source cache simulator.
Experiments run on benchmark datasets show that MUSTACHE outperforms the best page replacement heuristic (i.e., exact LRU), improving the cache hit ratio by $1.9\%$ and reducing the number of reads/writes required to handle cache misses by $18.4\%$ and $10.3\%$.
\end{abstract}
]



\printAffiliationsAndNotice{\mlsysEqualContribution} 

\newcommand{\tuple}[1]{(#1)}
\newcommand{\vas}{\mathcal{V}}
\newcommand{\pages}{\mathcal{P}}
\newcommand{\mem}{\mathcal{M}}
\newcommand{\kpages}{\hat{\pages}_{t+1:t+k}}
\newcommand{\candpages}{\mathcal{C}}
\newcommand{\rand}{\pi_{\text{RANDOM}}}
\newcommand{\fifo}{\pi_{\text{FIFO}}}
\newcommand{\opt}{\pi_{\text{OPT}}}
\newcommand{\lru}{\pi_{\text{LRU}}}
\newcommand{\mustache}{\pi_{\text{MUSTACHE}}}
\newcommand{\R}{\mathbb{R}}
\newcommand{\inst}{\bm{x}}
\newcommand{\X}{{\bf X}}
\newcommand{\dataset}{\mathcal{D}}
\newcommand{\features}{\mathcal{X}}
\newcommand{\hidden}{\bm{h}}
\newcommand{\cell}{\bm{c}}
\newcommand{\igate}{\bm{i}}
\newcommand{\fgate}{\bm{f}}
\newcommand{\ogate}{\bm{o}}
\newcommand{\weights}{\bm{W}}
\newcommand{\bias}{\bm{b}}

\newcommand{\labels}{\mathcal{Y}}
\newcommand{\params}{\bm{\theta}}
\newcommand{\model}{m^*}
\newcommand{\loss}{\mathcal{L}}
\newcommand{\rec}{f_{\text{rec}}}
\newcommand{\dir}{f_{\text{dir}}}
\newcommand{\dirrec}{f_{\text{dir-rec}}}
\newcommand{\mimo}{f_{\text{MIMO}}}
\newcommand{\dirmimo}{f_{\text{dir-MIMO}}}
\newcommand{\arima}{\text{ARIMA}(p, d, q)}

\newcommand{\gabri}[1]{\todo[inline,color=red!60]{{\bf Gabri:} #1}}
\newcommand{\fabio}[1]{\todo[inline,color=violet!60]{{\bf Fabio:} #1}}
\newcommand{\lore}[1]{\todo[inline,color=green!60]{{\bf Lore:} #1}}

\section{Introduction}
\label{sec:introduction}

Caching is a well-known technique used to store data on temporary storage (i.e., a {\em cache}) so that they can be accessed faster when requested in the future.
A {\em cache hit} occurs whenever the requested data can be found in the cache; otherwise, a {\em cache miss} arises. 
Cache hits are served by reading data from the cache, which is faster than recomputing a result or reading it from slower storage; the more requests that can be served from the cache, the faster the system performs.

Generally speaking, a cache can be found at each level of the memory hierarchy.
Indeed, several caches may exist between the CPU and main memory, and they are usually managed entirely by hardware. For example, the {\em translation lookaside buffer} (TLB), which is part of the {\em memory management unit} (MMU), stores the recent mappings of virtual to physical memory addresses. Doing so makes the entire address translation process quicker than if the CPU had to access the whole {\em page table} stored in the main memory.

When implementing {\em virtual memory}, the operating system (OS) uses the main memory (RAM) as a cache for secondary storage (disk). 
In this work, we consider a generic memory cache, possibly managed by the OS kernel, that stores a subset of referenced memory pages.


As this memory cache gets populated with pages referenced by active working sets (e.g., via {\em demand paging}), it may eventually be filled up.
Therefore, if a running task requests a page that is not in the cache, the OS must cleverly decide which page to evict from the cache to make room for the newly referenced page while maintaining a high cache hit ratio, i.e., a high percentage of cache hits.

Theoretically, the optimal cache replacement algorithm that maximizes the hit ratio (also known as OPT or Bélády's optimal page replacement policy) is known and works as follows: when a page needs to be swapped in from a slower yet larger memory (e.g., disk) to a quicker yet smaller memory cache (e.g., RAM), the OS swaps out the page in the cache whose next use will occur farthest away in the future.
Of course, such a policy cannot be implemented in practice as it would require the OS to know in advance what page(s) will be accessed later.
Thus, OSs typically use suboptimal, fixed cache eviction heuristics.
For example, assuming that the past is a good predictor of the future, many OSs swap out from the cache pages that are {\em least recently used} (LRU), implementing some LRU approximation, for instance, the ``second chance'' clock algorithm~\cite{corbato1969clock}.
Hence, existing cache eviction policies are generally effective with workloads that exhibit temporal/spatial locality. Still, they may fail to capture ``unconventional'' patterns of references, thereby losing the advantage of caching and, ultimately, causing the system to degrade.

To overcome this limitation, we propose a policy based on {\em \textbf{MU}lti-\textbf{ST}ep-\textbf{A}head Predictions for \textbf{C}ac\textbf{H}e \textbf{E}viction} (MUSTACHE). 
This is a new page replacement algorithm whose logic is {\em learned} from observed memory access requests rather than fixed as existing methods.
More specifically, we formulate the page request prediction problem as a categorical -- i.e., discrete-valued -- time series forecasting task.
Then, our method queries the learned page request forecaster to obtain the next $k$ predicted page memory references to better approximate the optimal OPT strategy.
As per how to implement the multi-step-ahead page request forecaster, several techniques can be used.

In this work, we only consider deep learning techniques, both standard and specifically designed for handling sequence data in general and time series in particular.
Among standard approaches, we use multilayer perceptron (MLP). In addition, we study architectures built upon recurrent neural networks (RNNs), such as ``vanilla'' long short-term memory networks (LSTM~\cite{hochreiter1997nc}) and sequence-to-sequence (seq2seq) models like Encoder-Decoder LSTM~\cite{sutskever2014neurips}. 

To train any of these methods, we first collect a very large dataset that logs the memory traces of the execution of benchmark programs contained in the Princeton Application Repository for Shared-Memory Computers (PARSEC).\footnote{\url{https://parsec.cs.princeton.edu/index.htm}}

We evaluate the prediction accuracy of each learned page request forecaster offline using a test set previously held out from the entire dataset above. 
Thus, we consider the best-performing forecaster as the building block of our MUSTACHE policy and integrate it into an open-source cache simulator.\footnote{\url{https://github.com/JoeBalduz/Page-Replacement-Memory-Simulator}}
Finally, we compare MUSTACHE with existing page cache replacement baselines already implemented in the cache simulator when running the test portions of all the programs in the PARSEC suite.
Experiments demonstrate the superiority of our approach, as MUSTACHE improves the cache hit ratio by $1.9\%$ and reduces the number of disk operations (i.e., number of reads/writes) by $18.4\%$ and $10.3\%$ compared to the best heuristic (i.e., exact LRU).

To summarize, we provide the following contributions:
\begin{enumerate}
    \item[{\em(i)}] We collect a very large dataset containing the memory accesses referenced during the execution of PARSEC benchmark programs;
    \item[{\em(ii)}] We frame the problem of page request prediction into a multi-step-ahead time series forecasting task;
    \item[{\em(iii)}] We train several deep learning models for page request prediction;
    \item[{\em(iv)}] We implement our cache replacement strategy (MUSTACHE) using the best-performing page request predictor learned, and we prove its superiority over well-known baselines;
    \item[{\em(v)}] We will publicly release all the data and implementation code of MUSTACHE along with our experiments.\footnote{\url{https://anonymous.4open.science/r/MUSTACHE/}} 
    \item[{\em(vi)}] We provide a detailed discussion on the feasibility and limitations of our proposed method if integrated into a real-world system. 
\end{enumerate}

The remainder of this paper is organized as follows.
We review related work in Section~\ref{sec:related_work}.
Section~\ref{sec:background} contains useful background concepts.
Section~\ref{sec:method} describes MUSTACHE, our proposed cache eviction policy, whereas its implementation is detailed in Section~\ref{sec:implementation}.
In Section~\ref{sec:experiments}, we validate our method, and possible limitations of it are discussed in Section~\ref{sec:limitations}.
Finally, Section~\ref{sec:conclusion} concludes the paper.
\section{Related Work}
\label{sec:related_work}

The need for a novel approach to OS design that leverages ML/AI techniques rather than relying exclusively on human wisdom has been advocated in~\cite{yiying2019sigops}.
There, the authors envision a new ``learned'' OS paradigm, where ML/AI can support at least three types of traditional OS components: {\em dynamic configurations} (e.g., a smarter timer interrupt settings rather than fixed apriori), {\em policy generation} (e.g., cache eviction policies learned from past memory references as opposed to heuristics), and {\em mechanism implementation} (e.g., ML-based mapping of virtual to physical addresses instead of fixed page table mapping).
However, the authors discuss the opportunities and challenges of this new OS design approach in general, without proposing any concrete solution to a specific problem. 

In the following, we review the main ML/AI techniques proposed in the literature to improve memory prefetching.

In~\cite{jain2016isca}, the authors apply Bélády's algorithm to past memory accesses and use this knowledge to learn future cache replacement decisions. The method has been evaluated using SPEC 2006 CPU benchmarks, showing better performance over LRU.

In~\cite{hashemi2018icml}, memory prefetching is seen as an $n$-gram model in natural language processing. 
Specifically, the authors show how recurrent neural networks can serve as a drop-in replacement.
Similarly, in~\cite{braun2019aidarc}, the authors adopt an LSTM neural network to learn memory access patterns by training individual models on microbenchmarks with well-characterized patterns of memory requests. Following this direction,~\cite{peled2019taco} proposes a context-based neural network prefetcher that dynamically adapts to arbitrary memory access patterns. 
In particular, it correlates program and machine contextual information with memory access patterns, using online training to identify and dynamically adjust to unique access patterns exhibited by the code. 
In this way, the prefetcher can discern the useful context attributes and learn to predict previously undetected access patterns, even within noisy memory access streams, by targeting semantic locality.

In~\cite{ayers2020asplos}, the authors introduce a novel methodology to classify the memory access patterns of applications. The proposed approach leverages instruction dataflow information to uncover a wide range of access patterns and their combinations (prefetch kernels), such as reuse, strides, reference locality, and complex address generation. These kernels are then used to compute the next address for most top-missing instructions.
Another deep learning methodology, called Voyager, is presented in~\cite{shi2021asplos}. Voyager learns delta and address correlations thanks to its hierarchical structure that separates addresses into pages and offsets. It introduces a mechanism for learning important relations among pages and offsets.

In this paper, to the best of our knowledge, we are the first to deal with ML/AI-empowered caching. Although related, caching and prefetching are two different techniques to reduce storage access times~\cite{patterson1994pdis}. 
As a matter of fact, a central issue in any prefetching strategy is the interaction with the activities of page cache replacement~\cite{albers2000jacm,kaplan2002ismm}. 
\section{Background}
\label{sec:background}
This section formulates two well-known problems useful to understand our proposed method: {\em page replacement policies} and {\em time series forecasting}.
\subsection{Page Replacement Policies}
\label{subsec:page_replacement}
We consider a multiprogramming OS, where each process can address $N$ virtual memory locations $\vas=\{v_0, \ldots, v_{N-1}\}$.
Moreover, this virtual address space is divided into a set of $L$ fixed-size {\em logical} pages $\pages=\{p_0, \ldots, p_{L-1}\}$, i.e., $\vas = \bigcup_{i=0}^{L-1}p_i$.
Each page $p_i$ is a contiguous sequence of $B$ memory addresses, namely $|p_i| = B~\forall i=0,\ldots, L-1$, therefore $N = L * B$.
Furthermore, the OS uses a memory cache $\mem$ to store a subset of the whole virtual address space $\vas$. 
More specifically, $\mem$ is itself divided into $K$ fixed-size physical page frames, and at each point in time $t$\footnote{A more accurate definition of time granularity is given in Section~\ref{subsec:mustache}.} it contains a subset $\pages_t$ of all the pages in $\pages$, namely $\pages_t\subset \pages$, where $|\pages_t|\leq |\mem| = K$ and generally $K \ll L$. 
The remaining set $\pages\setminus \pages_t$ of pages can instead be stored on a slower secondary memory $\mem'$.
At most, only a fraction $K/L$ of all the addressable pages are stored in the faster cache $\mem$.

Whenever a task running on the system references a page frame $p$ during its execution at a specific time $t$, only one of the following two events may occur: {\em (i)} $p\in \pages_t$ ({\em cache hit}), the page is already in the cache $\mem$ and the request can be served straight away, or {\em (ii)} $p\notin \pages_t$ ({\em cache miss}\footnote{This may be referred to as {\em page fault} in this context.}), the page must be first loaded from the secondary storage $\mem'$ to $\mem$ before the request being served.
In the latter case, as long as $\mem$ is not full, i.e., until all the $K$ frames are allocated, the requested page $p$ can be painlessly loaded from $\mem'$ into one of the free slots available in $\mem$, and the new set of pages stored in the cache becomes $\pages_{t+1} = \pages_t \cup \{p\}$.
On the other hand, if $\mem$ is full upon a cache miss, the OS must first make room for the newly referenced page $p$ before it can load it from $\mem'$. 
In other words, the OS must pick one of the frames $p^*\in \pages_t$ currently stored in the cache, swap $p^*$ out to $\mem'$, and finally swap $p$ into $\mem$.
Thus, the new set of pages stored in $\mem$ becomes $\pages_{t+1} = \pages_t \setminus \{p^*\} \cup \{p\}$.

To select the page $p^*\in \pages_t$ that will be replaced by the new $p$, the OS relies on a cache eviction policy $\pi$.
More formally, $\pi:\{\pages_t\} \mapsto \pages_t$ is a {\em choice function} that maps the whole set of pages $\pages_t$ to one of its elements $p^* \in \pages_t$.
For example, a straightforward policy would be to uniformly select at random one of the pages currently stored in $\mem$, i.e., $\rand:~p^* \in_R \pages_t$.
Other well-known page replacement strategies are: $\fifo$, which selects $p^*$ as the page that was firstly loaded in $\mem$ (i.e., the ``oldest'' page in $\mem$); $\lru$, which chooses $p^*$ as the page that was least recently referenced; $\opt$, which is the optimal yet not directly implementable policy, removes the page $p^*$ as the farthest one requested in the future.

In this work, we propose a new page replacement policy $\mustache$ that is {\em learned} from historical memory access requests rather than fixed apriori.
More specifically, $\mustache$ relies on solving a categorical time series forecasting problem, whose general definition is given below.

\subsection{Time Series Forecasting}
\label{subsec:time_series_forecasting}

Let $\{Y_t,~t\in T\}$ be a time series process, namely a stochastic process represented by a collection of random variables indexed by time $\{Y_t\}$. 
In the following, we assume the index set $T$ is countably infinite, e.g., $T \subseteq \mathbb{Z}$, and we refer to $\{Y_t\}$ as a discrete-time stochastic process.\footnote{The theory can be extended to continuous time, but in this work we assume data are observed at discrete, equally-spaced time intervals.}
We define a time series $\{y_t,~t=1,\ldots, n\}$, or simply $y_{1:n}$, as a realization of $\{Y_t\}$, i.e., a finite sample of $n$ observations of the random variables underlying the discrete-time stochastic process collected at equally-spaced points in time, namely the $n$-dimensional random vector $\bm{y} = (Y_1=y_1, \ldots, Y_n=y_n)$.
More specifically, we consider a univariate time series, where at each time $t$ the observation $y_t$ is a single, scalar value, e.g., $y_t\in \mathbb{R}$.\footnote{This can be generalized to multivariate time series, where each $\bm{y}_t\in \mathbb{R}^d$ is a $d$-dimensional vector.}


Suppose we have access to a univariate time series $\bm{y} = \{y_{t-w+1}, \ldots, y_{t-1}, y_t, y_{t+1}, \ldots, y_{t+k}\}$. 
We consider the standard multi-step-ahead forecasting problem as follows. 
Let $y_{t-i},~i=0,\ldots,w-1$ be the sequence of past $w$ observations up to time $t$.
Intuitively, the goal of multi-step-ahead forecasting is to estimate the future $k$ observations $y_{t+j},~j\in \{1,\ldots,k\}$, denoted by $\hat{y}_{t+j}$, leveraging the past history $y_{t-w+1:t}$ along with possibly other signals.
More formally, in its most generic description, the multi-step-ahead forecasting problem resort to finding a predictive model $f$, such that:
\begin{equation}
\label{eq:msa-forecasting-1}
\hat{y}_{t+j} = f(y_{t-w+1:t}, \bm{x}_{t-w+1:t}, \bm{u}_{t-w+1:t+j}, \bm{s},j;\params),
\end{equation}
where:
\begin{itemize}
    \item $\hat{y}_{t+j}$ is the forecast after $j$ time steps output by the model $f$;
    \item $y_{t-w+1:t} = (y_{t-w+1}, \ldots, y_t)$ are the observations of the target (also referred to as {\em endogenous} inputs) over a look-back window $w$;
    \item $\bm{x}_{t-w+1:t} = (\bm{x}_{t-w+1}, \ldots, \bm{x}_t)$ are external, time-dependent predictors, i.e., {\em exogenous} inputs, again measured over the same look-back window $w$;
    \item $\bm{u}_{t-w+1:t+j} = (\bm{u}_{t-w+1}, \ldots, \bm{u}_{t+j})$ are known future inputs across the whole time horizon (e.g., date information, such as the day-of-week or month);
    \item $\bm{s}$ is some static metadata that does not depend on time;
    \item $\params$ is the overall vector of {\em parameters} of the forecasting model. Despite some ambiguity on the terminology exists,\footnote{\url{https://en.wikipedia.org/wiki/Nonparametric\_statistics\#Definitions}} in this work we only consider {\em parametric} models, namely models whose structure is assumed fixed, independently on the number of training observations.
\end{itemize}
We can rewrite (\ref{eq:msa-forecasting-1}) more concisely as:
\begin{equation}
\label{eq:msa-forecasting-2}
\hat{y}_{t+j} = f(y_{t-w+1:t}, \bm{z}_{t,w}, j;\params),
\end{equation}
where $\bm{z}_{t,w}$ concatenates all non-endogenous inputs.

Notice that the above formulation is flexible enough to express basic forecasting models, where predictions of future target values are obtained only from its past observations (i.e., endogenous inputs), namely $\hat{y}_{t+j} = f(y_{t-w+1:t};\params)$.
Furthermore, one-step-ahead forecasting is just a special case of multi-step-ahead formulation, where $j=1$.

Several approaches have been proposed in the literature to solve the multi-step-ahead forecasting problem, i.e., to find the best $f$ above.
In Section~\ref{subsec:strategies}, we discuss the time series forecasting methods used in this work for implementing our new page replacement policy $\mustache$.
\section{Proposed Method}
\label{sec:method}

\subsection{The MUSTACHE Replacement Policy}
\label{subsec:mustache}
We consider the memory page requests spawn by the generic active workload on a system as a discrete-time stochastic process $\{Y_t\}$; any realization of this process is a {\em categorical} univariate time series $\{y_t,~t=1,\ldots,n\}$, where each observation $y_t$ is a discrete value, i.e., $y_t \in \pages$. 
We assume that observations are collected at equally-spaced points in time, namely page requests are generated at regular intervals (e.g., at every CPU clock cycle).

Moreover, suppose there exists a $k$-step-ahead forecasting model $f$ that, at any time $t$, given the last $w$ page requests $y_{t-w+1:t}$ along with additional inputs $\bm{z}_{t,w}$, is able to predict the next $k$ page references $\hat{y}_{t+1:t+k}$, according to (\ref{eq:msa-forecasting-2}).

Let $\pages_t$ be the set of pages stored in the memory cache $\mem$ at time $t$, and assume $\mem$ is full (i.e., $|\pages_t| = |\mem| = K$). Furthermore, $y_t$ generates a page fault (i.e., a cache miss), namely $y_t = p \notin \pages_t$.
To serve this request, we propose the system uses the following page replacement policy, called $\mustache$. 

First, it queries the $k$-step-ahead forecasting model $f$ in order to retrieve the set of $k$ predicted future page references $\hat{y}_{t+1:t+k}$: let us call this set $\kpages \subset \pages$, where $|\kpages| \le k$.\footnote{Notice that the same page may appear multiple times in the sequence of $k$ predicted references.}
Then, it computes the intersection between the set of pages currently stored in the cache and the collection of predicted referenced pages, i.e., $\pages_t \cap \kpages$.
It is worth remarking that accurately predicting the next $k$ page requests would allow us to better approximate the optimal replacement algorithm.
Intuitively, the page $p^*$ to be evicted must be picked from the set of {\em candidates} $\candpages_t$ containing all the pages that are in the cache, except those in the intersection with the predictions, i.e., $\candpages_t = \pages_t \setminus (\pages_t \cap \kpages)$. 
Thus, we distinguish between two cases: {\em(i)} $\candpages_t = \pages_t$, or {\em(ii)} $\candpages_t \subset \pages_t$.
The former {\em(i)} means that none of the predicted pages referenced in the future are currently in the cache (i.e., $\pages_t \cap \kpages = \emptyset$), thereby -- as far as the OS is {\em currently} concerned -- all of them will generate a cache miss. 
In this case, predictions provided by $f$ do not help, as they do not restrict the space of candidates to eviction, and $\mustache$ may fall back to one of the existing page replacement policies, e.g., $\lru$.
The latter {\em(ii)}, instead, implies that at least one of the pages that are already in the cache will be requested within the next $k$ accesses, according to $f$ (i.e., $\pages_t \cap \kpages \neq \emptyset$).
In this case, we further consider two events, separately: {\em(ii.a)} $\candpages_t = \emptyset$, or {\em(ii.b)} $\candpages_t \neq \emptyset$. 
The first occurs when the intersection between the set of pages currently stored in $\mem$ and the pages predicted to be accessed in the future is precisely equivalent to the former, i.e., $\pages_t \cap \kpages = \pages_t$.
Coherently, $\candpages_t =\emptyset$, since no actual page in $\mem$ is a clear optimal candidate for replacement, as all of them are predicted to be referenced.
Therefore, to break the tie, $\mustache$ may decide to evict the page that will probably be referenced farthest away in the future, according to the forecast.
However, {\em(ii.a)} is an improbable event, mainly because $f$ would require generating predictions over a horizon at least as large as the size of $\mem$, whereas, usually, $k < K$.
On the other hand, when {\em(ii.b)} occurs, {\em any} page $\widetilde{p}$ that is stored in $\mem$ (i.e., $\widetilde{p} \in \pages_t$) but is not part of the forecast (i.e., $\widetilde{p} \notin \kpages$) is a candidate for replacement. 
Except for the edge case where $|\pages_t \cap \kpages| = K-1$, and thus there is only one page candidate to evict, in every other case, the page $p^*$ to be replaced can be chosen using any well-known strategy (e.g., again $\lru$) amongst the set of candidates $\candpages_t$.

\begin{figure}[h]
\centering
\includegraphics[width=\columnwidth]{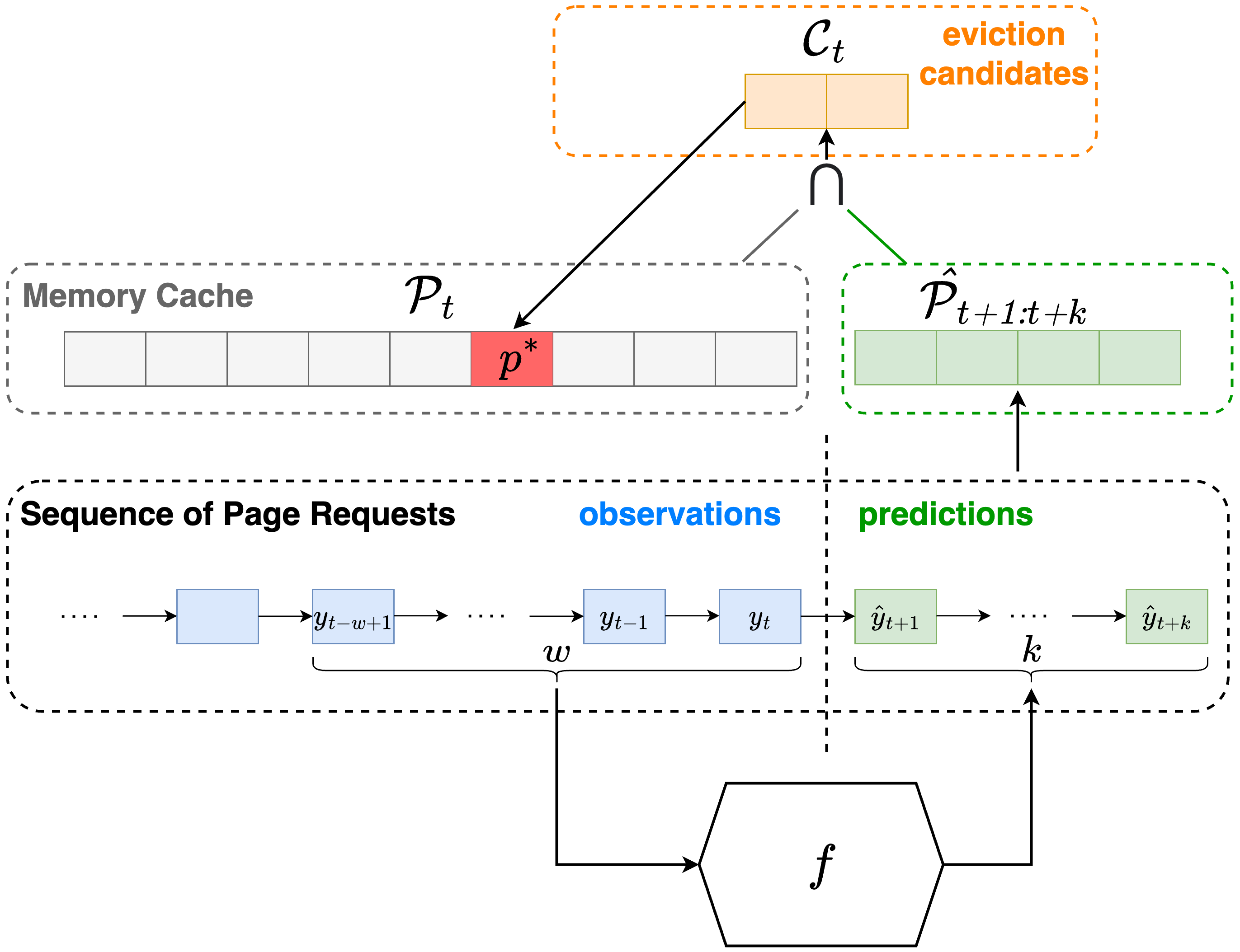}
\caption{Overview of our MUSTACHE page replacement policy.}
\label{fig:framework}
\end{figure}

An overview of our MUSTACHE page replacement policy is depicted in Fig.~\ref{fig:framework}.

\subsection{The MUSTACHE Algorithm}
\label{subsec:algorithm}
The pseudocode of the MUSTACHE replacement policy ($\mustache$)\footnote{In the following, we refer to the name of each policy directly (MUSTACHE instead of $\mustache$, LRU instead of $\lru$, etc.)} described above is provided in Algorithm~\ref{alg:mustache}. 
Notice that we use LRU as the fallback page replacement strategy; however, any other well-known policy can be easily plugged in.
Moreover, with a slight abuse of notation, we assume the function \texttt{get\_farthest} takes the set of page access predictions output by $f$ and returns the page which will be referenced farthest away in the future, amongst those predicted pages.
To clarify how this works, consider, for example, that $f$ predicts the following {\em sequence} of $k=8$ page id requests: $(25, 19, 19, 42, 25, 37, 42, 19)$. 
Hence, the {\em set} of predicted pages $\hat{\pages}_{t+1:t+k} = \{25, 19, 42, 37\}$, as some of the pages are repeated in the sequence of request forecasts.
Eventually, the page id returned by \texttt{get\_farthest} is $37$, as that is the page whose first occurrence happens at the latest point of the sequence of predictions, according to $f$.

\begin{algorithm}[ht]
\caption{\texttt{MUSTACHE} Page Replacement}
\label{alg:mustache}
\begin{algorithmic}[1]
\STATE \textbf{Input:}\\
$\pages_t$ \hfill\COMMENT{The set of pages stored in the cache $\mem$ at time $t$}\\
$y_{t-w+1:t}$ \hfill\COMMENT{The last $w$ page requests observed}\\
${\bm z}_{t,w}$ \hfill\COMMENT{Additional input signals}\\
$f$ \hfill\COMMENT{The page request forecaster}
\STATE \textbf{Output:}\\
$p^*\in \pages_t$ \hfill\COMMENT{The page to be replaced from the cache $\mem$}
\item[]
\STATE {\bf function }{\texttt{\textbf{MUSTACHE}}($\pages_t, y_{t-w+1:t}, {\bm z}_{t,w}, f$):}
\STATE $\hat{\pages}_{t+1:t+k} \gets f(y_t, {\bm z}_{t,w})$ \hfill\COMMENT{The next $k$ predicted pages}
\STATE $\candpages_t \gets \pages_t \setminus (\pages_t \cap \kpages)$ \hfill\COMMENT{The set of candidates}
\IF{$\candpages_t = \pages_t$}
\STATE $p^* \gets \texttt{LRU}(\pages_t)$
\ELSE 
\IF{$\candpages_t = \emptyset$}
\STATE $p^* \gets \texttt{get\_farthest}(\hat{\pages}_{t+1:t+k})$
\ELSE 
\STATE $p^* \gets \texttt{LRU}(\pages_t \setminus \hat{\pages}_{t+1:t+k})$
\ENDIF
\ENDIF
\STATE \textbf{return} $p^*$
\STATE {\bf end function}
\end{algorithmic}
\end{algorithm}

Assuming the page request forecaster $f$ is already trained, the computational complexity of Algorithm~\ref{alg:mustache} boils down to the time required for accomplishing three main tasks: {\em (i)} calculating the predictions with $f$ (line 9); {\em (ii)} calling the fallback replacement strategy, e.g., LRU (lines 12 and 17); and {\em (iii)} computing the \texttt{get\_farthest} function.
Concerning {\em (i)}, this of course depends on how $f$ is implemented. For instance, if $f$ represents a (trained) deep neural network, predictions at inference time are computed with a constant number of sums of products of the input and possibly a final activation function. Overall, the time complexity of this step is linear in the input size, i.e., $O(d)$, where $d = dim(y_t \oplus {\bm z}_{t,w})$.
\section{Page Request Forecasting}
\label{sec:implementation}
In this section, we clarify how page request prediction can be framed as a multi-step-ahead time series forecasting problem. Furthermore, we describe different approaches to implement the page request forecaster $f$, which MUSTACHE uses to select the page that will be evicted from the memory cache.

\subsection{Categorical Time Series Forecasting}
\label{subsec:considerations}
When we introduced the general (multi-step-ahead) time series forecasting problem in Section~\ref{subsec:time_series_forecasting} above, we assumed that observations and output responses to predict are continuous, real values, i.e., $y_{t}\in \R$.
Instead, in the context of page request forecasting, these values are discrete, i.e., $y_{t}\in \pages$.
Therefore, we must frame the problem of page request prediction as a categorical time series forecasting task, where the output range of $f$ is a discrete set of pages.
In other words, we move from a regression to a multi-class classification task, where the number of pages determines the number of classes.
Somehow, this resembles the well-known next-word prediction problem~\cite{bengio2003jmlr} in natural language processing (NLP), where -- given a sequence of words as input (respectively, a sequence of page requests) -- the goal is to predict the word that will most likely occur next in the sequence (respectively, the page that will be accessed in the future).

As already highlighted in~\cite{hashemi2018icml}, however, a significant concern quickly becomes evident: the virtual address space $\vas$ of a process is extremely large, and -- to a lesser extent -- so does the range of pages $\pages$. 
For example, if $|\vas| = N = 2^{64}$ addressable bytes\footnote{Typically, 64-bit systems do not support full 64-bit virtual memory addresses (e.g., x86-64 and ARMv8 use only 48 bits).} and each page is $2^{12}=4,096$ bytes long, the range of predictions that $f$ must cover is still huge, i.e., $|\pages| = L = 2^{52}\approx 4.5$ quadrillion pages.
Thus, we cannot consider the whole set of pages $\pages$ as the label space since this would turn $f$ into an impracticable $L$-class classifier, which has to estimate a multinomial distribution over $L$ pages.
It turns out that some quantization mechanism is needed.

To achieve that, we can again take inspiration from NLP and restrict ourselves to a more manageable set of output labels by fixing a {\em vocabulary} of the most commonly referenced pages.
Luckily, programs tend to obey locality principles during their execution, i.e., only a relatively small (although still large in absolute numbers) and consistent set of pages are referenced.\footnote{According to the well-known 90/10 rule, 90\% of memory accesses span only 10\% of all the available addresses.}
The sparseness of page requests suggests that the adequate vocabulary size may be significantly smaller than the original set of pages $\pages$.
Other quantization mechanisms can also be designed to reduce the dimensionality of the label space (e.g., clustering pages that tend to be referenced closely together), but this is outside the main scope of this work.

Moreover, due to dynamic side-effects such as address space layout randomization (ASLR), different runs of the same program will lead to different raw memory references~\cite{spengler2003pax}. 
However, for a given layout, the program will behave consistently. 
Therefore, one potential strategy is to predict deltas rather than raw page references, i.e.,  $\delta_{t+j} = y_{t+j} - y_{t}$, as proposed by~\cite{hashemi2018icml}.
\begin{figure}[h]
\centering
\includegraphics[width=\columnwidth]{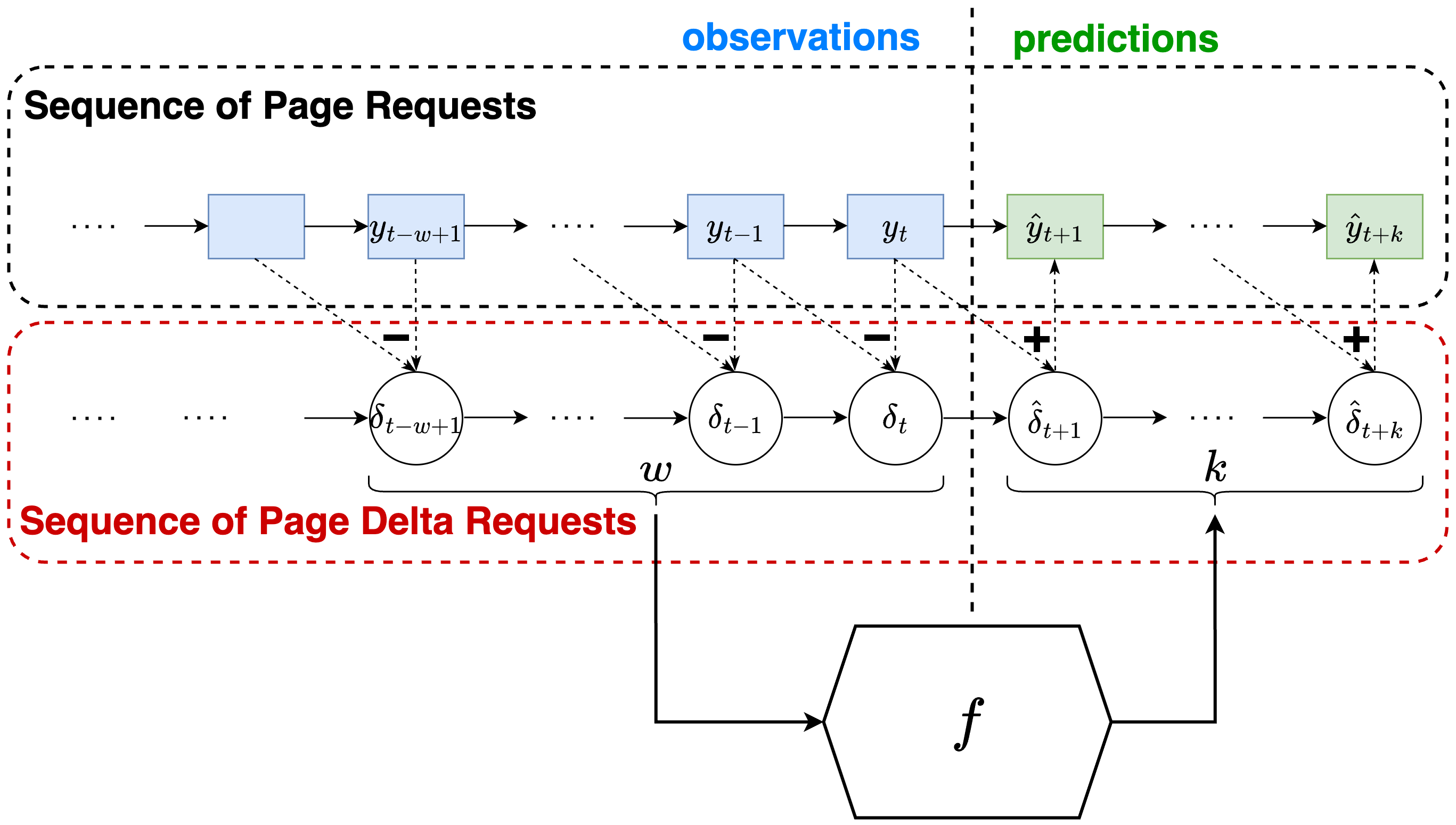}
\caption{Multi-step-ahead page delta forecasting.}
\label{fig:page-deltas}
\end{figure}

These will remain consistent across program executions and come with the benefit that the number of uniquely occurring deltas is often orders of magnitude smaller than uniquely referenced page addresses.
To clarify how this works, consider the following sequence of page requests: $(y_t=73, y_{t+1}=81, y_{t+2}=67, y_{t+3}=67, y_{t+4}=75)$. 
This will be transformed into the sequence of page deltas $(\delta_{t+1}=8, \delta_{t+2}=14, \delta_{t+3}=0, \delta_{t+4}=8)$.
In our models, we use a sequence of page deltas as inputs instead of raw page addresses, as shown in Fig.~\ref{fig:page-deltas}.

\subsection{Strategies for Page (Delta) Request Forecasting}
\label{subsec:strategies}

Previously, we discussed how to transform a timely-ordered sequence of page requests into a categorical time series of page deltas to reduce the size of the output label set.
Therefore, any method for time series forecasting can be used to predict the next $k$ page deltas referenced (i.e., our observations $y_{t+j}$ becomes, in fact, $\delta_{t+j} = y_{t+j} -y_t$). 
It is worth remarking that this approach does not limit our method, as the original, raw page references can always be obtained by adding the predicted page delta to the page value observed at the previous time step.

Generally speaking, time series forecasting approaches can be broadly categorized into statistical methods (e.g., ARIMA), standard machine learning techniques (e.g., XGBoost), deep learning solutions based on standard feed-forward neural networks (e.g., MLP), convolutional or recurrent neural networks (e.g., LSTM). 
For a comprehensive survey on
(multi-step-ahead) time series prediction, we invite the reader to refer to~\cite{liu2021access,lim2021rs,chandra2021access}.
\begin{figure*}[h!]
     \centering
     \begin{subfigure}[b]{0.25\textwidth}
         \centering
         \includegraphics[width=\textwidth]{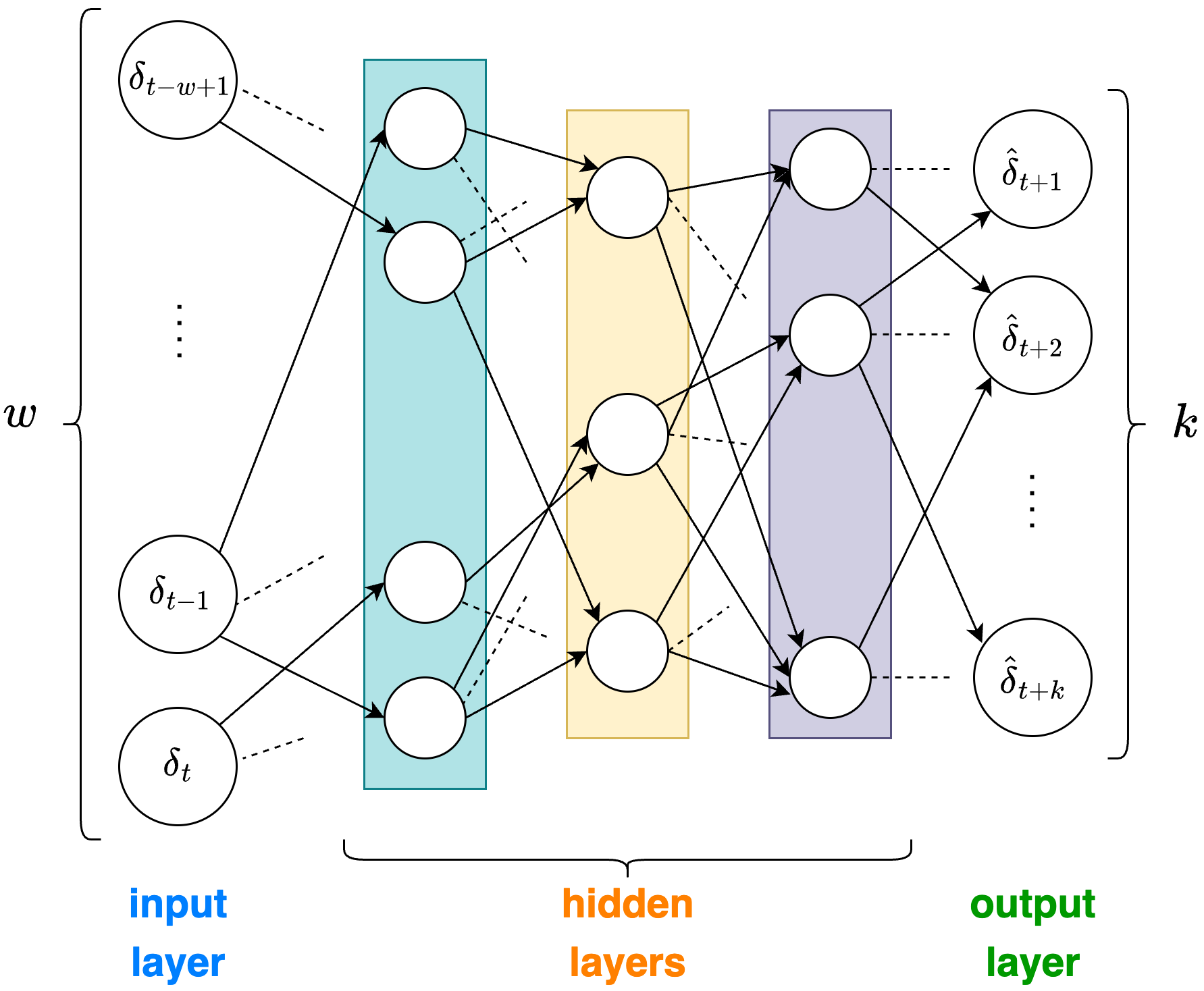}
         \caption{Multilayer Perceptron (MLP).}
         \label{fig:mlp}
     \end{subfigure}
      \begin{subfigure}[b]{0.37\textwidth}
         \centering
         \includegraphics[width=\textwidth]{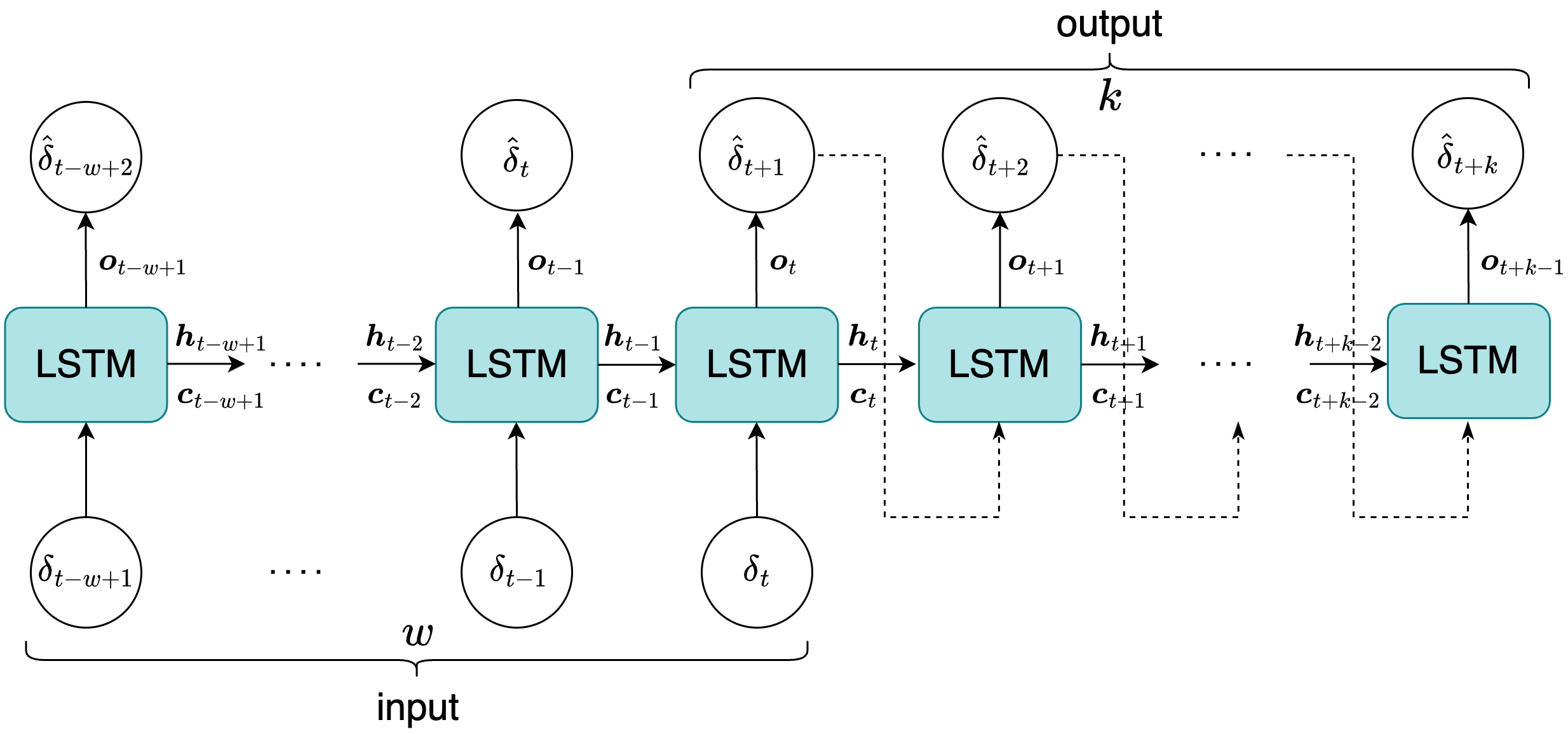}
         \caption{``Vanilla'' LSTM.}
         \label{fig:lstm}
     \end{subfigure}
     \begin{subfigure}[b]{0.37\textwidth}
         \centering
         \includegraphics[width=\textwidth]{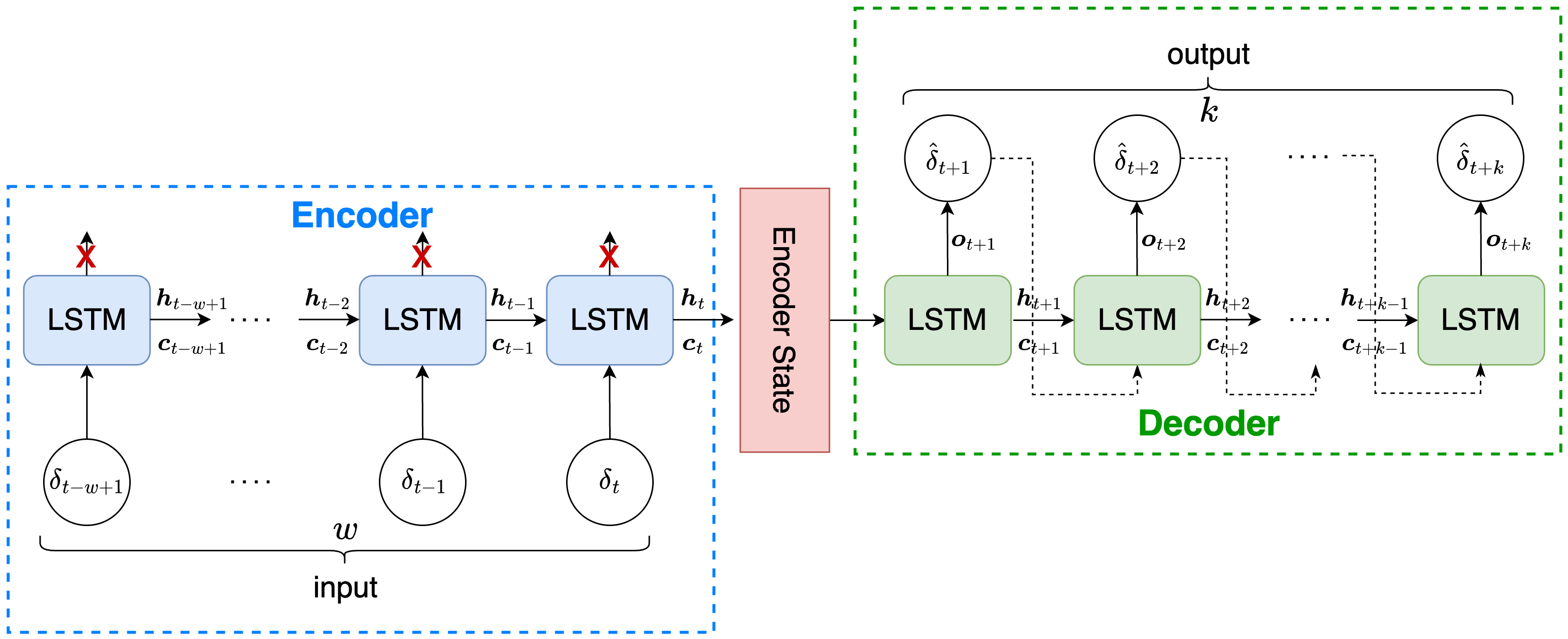}
         \caption{Encoder-Decoder LSTM (ED-LSTM).}
         \label{fig:ed-lstm}
     \end{subfigure}
        \caption{Multi-step-ahead page delta request forecasters used by MUSTACHE.}
        \label{fig:page-forecasters}
\end{figure*}
We consider cutting-edge deep learning methods in this work as they achieve state-of-the-art performance in many time series forecasting tasks. 
Specifically, we examine standard multilayer perceptron (MLP), ``vanilla'' LSTM, and Encoder-Decoder LSTM (ED-LSTM), whose architectures are depicted in Fig.~\ref{fig:page-forecasters}. 

\section{Experiments}
\label{sec:experiments}
To validate our newly proposed page replacement policy MUSTACHE, we perform the following steps:
\begin{enumerate}
    \item We collect a very large dataset of memory traces from the execution of the programs contained in the PARSEC benchmark suite;
    \item We train three page request forecasters using deep learning techniques and test their offline accuracy;
    \item We consider the best-performing page request forecaster as $f$ in our MUSTACHE page replacement algorithm (see Algorithm~\ref{alg:mustache});
    \item We use a publicly available, open-source page replacement simulator to compare the performance of our MUSTACHE policy against well-known baselines (Random, FIFO, LRU, CLOCK, and OPT) using standard metrics (i.e., cache hit ratio and number of I/O operations to handle cache misses). Furthermore, we compare MUSTACHE against the Adaptive Cache Replacement (ARC) baseline~\cite{megiddo2003fast} implemented by another open-source simulator.
\end{enumerate}

\subsection{Dataset Collection}
\label{subsec:dataset}
To create a suitable training set for learning a page request forecaster, we use the Intel Pin toolkit,\footnote{\url{https://www.intel.com/content/www/us/en/developer/articles/tool/pin-a-dynamic-binary-instrumentation-tool.html}} which is a dynamic binary instrumentation framework for the IA-32, x86-64, and MIC instruction-set architectures that, amongst other things, can track memory address requests.

We profile byte-level memory references of all the 19 programs contained in the PARSEC benchmark suite with Pin when executed on an Intel\textregistered Core\textsuperscript{TM} i7-9700 up to 4,70 GHz with 12 MB cache and 32 GB RAM.
It is worth noticing that the PARSEC benchmark programs cover several different areas, such as computer vision, video encoding, financial analytics, animation physics, and image processing. 
\begin{table}[ht]
\centering
\scalebox{0.72}{
\begin{tabular}{|c|c|c|c|c|}
\hline
\texttt{PC}        & \texttt{OP} & \texttt{MEM} & \texttt{N\_BYTES} & \texttt{MEM\_PREF}\\ \hline
{\tt 0x7f89388a7f1d} & {\tt R} & {\tt 0x7f89388d9ea0} & 8 & {\tt 0x5}\\ \hline
\end{tabular}
}
\caption{\label{tab:trace}An example of a single memory trace logged by Pin.
}
\end{table}

Table~\ref{tab:trace} shows an example of a single memory trace logged by Pin. 
Each record contains the following fields: 
\begin{itemize}
    \item {\tt PC}: The memory address in the program counter;
    \item {\tt OP}: The operation performed ({\tt R} = Read / {\tt W} = Write);
    \item {\tt MEM}: The (byte-level) memory address referenced;
    \item {\tt N\_BYTES}: The number of bytes read or written;
    \item {\tt MEM\_PREF}: The prefetched memory address.
\end{itemize}
The entire time series of memory references generated by a program can be simply obtained by considering the sequence of address values contained in the {\tt PC} and {\tt MEM} fields of each record (for that program).
Overall, we obtain a collection of around 340 million memory references for all the programs in the PARSEC suite. 
We use 90\% of the time series associated with each program for training each page request forecaster and the remaining 10\% for testing.

\subsection{Dataset Preprocessing}
\label{subsec:preprocessing}
First, we remove from each time series of memory access the leading subsequence preamble, which is common to all programs. 
In addition, we transform each sequence of byte-level memory references into a series of page-level requests for each program. 
Expressly, we assume a system whose fixed page size is $B = 2^{12}$ bytes. 
Moreover, to further reduce the dimensionality of our problem, we transform the sequence of the raw page referenced into a series of page deltas between any two consecutive requests.
To build our vocabulary of page deltas, we first remove the rarest one, i.e., those occurring only once in the training set.  
In the end, our vocabulary consists of approximately 5,100 page deltas.\footnote{The page deltas extracted from the training set include all the page deltas observed in the test set.}


\subsection{Training and Testing the Page Request Forecaster}
\label{subsec:training}
We train on the 90\% portion of each time series of page deltas the three multi-step-ahead page forecasting methods described in Section~\ref{sec:implementation} above: standard three-layer MLP, ``vanilla'' LSTM, and ED-LSTM. 
All the models are trained by minimizing categorical cross-entropy loss using Adam optimizer with a look-back window $w=100$. 
In Table~\ref{tab:models}, we summarize the main properties of all these trained models.
\begin{table}[ht]
\centering
\begin{tabular}{|c|c|}
\hline
\textbf{Model}        & \textbf{Hyperparameters}\\ \hline
MLP & \thead{\{{\tt \#params}$\approx315$k, {\tt batch\_size=}$256$,\\ {\tt \#epochs=}$15$, {\tt learning\_rate=}$10^{-4}$\}} \\ \hline
LSTM & \thead{\{{\tt \#params}$\approx20.5$M, {\tt batch\_size=}$256$,\\ {\tt \#epochs=}$15$, {\tt learning\_rate=}$10^{-4}$\}} \\ \hline
ED-LSTM & \thead{\{{\tt \#params}$\approx38.4$M, {\tt batch\_size=}$128$,\\ {\tt \#epochs=}$15$, {\tt learning\_rate=}$10^{-4}$\}} \\ \hline
\end{tabular}
\caption{\label{tab:models}Main properties of the trained models.
}
\end{table}


We validate all the trained page request forecasters by measuring the accuracy of their output predictions on the previously held out test set at a specific look-ahead horizon $k$.
More formally, let $(\delta_{t+1},\ldots,\delta_{t+k})$ be the $k$ actual page deltas observed after time $t$. 
Suppose that $(\hat{\delta}_{t+1}, \ldots, \hat{\delta}_{t+k})$ is the sequence of page deltas predicted by the forecaster $f$.
Therefore, we compute the $Accuracy@k$ of a model as the fraction of correct predictions appearing in the {\em right} order of the actual sequence of $k$ future page deltas. 
In other words:
\[
Accuracy@k = \frac{1}{k}\sum_{i=1}^k \mathbbm{1}(\delta_{t+i}=\hat{\delta}_{t+i}),
\]
where $\mathbbm{1}(\cdot)$ is the well-known 0-1 indicator function that evaluates to 1 if $\delta_{t+i}=\hat{\delta}_{t+i}$, or 0 otherwise.

In Table~\ref{tab:accuracy}, we show the $Accuracy@k$ of each model measured on the test set under different values of the forecasting horizon, i.e., $k=\{10, 20, 30\}$.
\begin{table}[ht]
\centering
\begin{tabular}{c|c|c|c|}
\cline{2-4}
& \multicolumn{3}{c|}{$Accuracy@k$}\\ 
\hline
\multicolumn{1}{|c|}{\textbf{Model}} & $k=10$ & $k=20$ & $k=30$\\ \hline
\multicolumn{1}{|c|}{MLP} & $0.68$ & $0.64$ & $0.61$\\ \hline
\multicolumn{1}{|c|}{LSTM} & ${\bf 0.87}$ & ${\bf 0.83}$ & ${\bf 0.80}$\\ \hline
\multicolumn{1}{|c|}{ED-LSTM} & $0.84$ & $0.77$ & $0.68$\\ \hline
\end{tabular}
\caption{\label{tab:accuracy}Test accuracy of each page request forecaster under different values of the forecasting horizon $k$.
}
\end{table}

The best-performing page delta request forecaster is the plain standard LSTM. Moreover, this method is more robust than competitors as the prediction horizon increases.
In the following, we use LSTM as the predictor $f$ described in Algorithm~\ref{alg:mustache} to implement our MUSTACHE page replacement policy. 

\subsection{Page Replacement Simulator}
\label{subsec:simulator}
To validate the effectiveness of our page replacement policy, we integrate MUSTACHE into an existing, open-source page replacement simulator.\footnote{\url{https://github.com/JoeBalduz/Page-Replacement-Memory-Simulator}}
We first refactor this tool, originally written in C, in Python for smoother integration with the predictive models developed with PyTorch.
The simulator implements five major page replacement policies, which we use as baselines: Random, FIFO, LRU (exact), CLOCK (an LRU approximation), and OPT (i.e., the optimal policy).
Random simply removes from the cache a page chosen uniformly at random.
FIFO evicts the ``oldest'' page in the cache (i.e., the page that first entered the cache amongst those currently stored).
Exact LRU removes the least recently used page, whereas CLOCK approximates true LRU utilizing a combination of FIFO and an array to keep track of the bits used to give the queued page a ``second chance'' before being selected for eviction. 
OPT is the provably optimal strategy that swaps out the page whose next use will occur farthest away in the future.
In addition, we use another open-source simulator\footnote{\url{https://gist.github.com/pior/da3b6268c40fa30c222f}} that implements the Adaptive Cache Replacement policy (ARC). ARC tries to improve the LRU strategy by splitting the cache into two lists for recently and frequently referenced entries.

We configure both page replacement simulators assuming 32-bit logical address (i.e., $2^{32}$ bytes virtual address space) and all (logical) pages and (physical) frames are $2^{12} = 4,096$ bytes long.
It turns out that the corresponding page table has $2^{32}/2^{12}$ entries, i.e., approximately 1 million entries. 
We consider a memory cache $\mem$ whose size is $K=40$KiB.

Hence, we run the page replacement simulators on the test portion of each of the 19 benchmark programs in the PARSEC suite we previously held out.

For each sequence, we measure the page replacement simulator's performance when implementing one of the following policies: Random, FIFO, LRU, CLOCK, ARC, OPT, and our MUSTACHE. 
Specifically, we consider two key metrics: the cache hit ratio, and the number of I/O operations (i.e., reads/writes) to handle cache misses.\footnote{The ARC simulator reports only the cache hit ratio.}

\subsection{Results}
\label{subsec:results}
In Table~\ref{tab:results}, we show the values of all the evaluation metrics considered for every page replacement policy. 
Specifically, we may observe that MUSTACHE ($k=30$) achieves the highest cache hit ratio of all the heuristic baselines except, of course, the optimal strategy (OPT).
At first sight, the improvement registered by MUSTACHE over the best baseline, i.e., exact LRU, ($1.9\%$) might seem bland; on the contrary, it is significant if we consider that: {\em(i)} the Random baseline already exhibits a pretty high cache hit ratio\footnote{It is well-known that random page replacement surprisingly good in practice.} and {\em(ii)} MUSTACHE halves the gap between the best baseline and OPT.
\begin{table}[ht]
\centering
\begin{tabular}{|c|c|c|c|}
\hline
\textbf{Policy} & \textbf{Hit Ratio} $\uparrow$ & \textbf{\#Reads} $\downarrow$ & \textbf{\#Writes} $\downarrow$\\ 
\hline
Random & $0.879$ & $1,240,957$ & $280,784$\\ \hline
FIFO & $0.884$ & $1,187,105$ & $288,842$\\ \hline
CLOCK & $0.898$ & $1,042,184$ & $227,550$\\ \hline
LRU & $0.908$ & $939,978$ & $176,706$\\ \hline
ARC & $0.906$ & N/A$^\dagger$ & N/A$^\dagger$ \\ \hline
MUSTACHE & ${\bf0.925}$ & ${\bf767,328}$ & ${\bf158,519}$\\ \hline\hline
OPT & ${0.945^*}$ & ${560,219^*}$ & ${115,022^*}$\\ \hline
\end{tabular}
\caption{\label{tab:results}Performance evaluation of all the page cache replacement policies ($^*$OPT is obviously unbeatable; $^\dagger$This information is not available for the ARC simulator).
}
\end{table}

Another crucial factor to consider when evaluating a page replacement algorithm is the number of I/O interactions (i.e., reads/writes) it requires upon cache misses. 
From Table~\ref{tab:results}, we can see that MUSTACHE again outperforms any other non-optimal competitors. Specifically, it reduces the number of reads and writes w.r.t. exact LRU by $18.4\%$ and $10.3\%$, respectively.
Unfortunately, the ARC simulator does not provide this information. 
However, we presume that ARC exhibits a trend in the number of I/O operations similar to that of LRU.
This result further testifies that MUSTACHE generates a lower number of page faults and, therefore, ``wastes'' fewer CPU cycles performing I/O operations due to page swapping to and from slower secondary storage.

\subsection{Ablation Study}
\label{subsec:ablation}
There are four fundamental hyperparameters of the page request forecaster $f$ that may impact the performance of our proposed MUSTACHE page replacement strategy: {\em (i)} the window size $w$ of past observations, {\em (ii)} the look-ahead $k$ of future predictions, and {\em (iii)} the size $v$ of the vocabulary of pages considered.
In addition, the cache size $K$ and the page size $B$ also plays a significant role in our experiments. 

Due to space limitation, in this work, we study the effect of several values of the prediction horizon $k$ on the first 1 million page requests of the test set. 
We leave a more comprehensive analysis, including also other parameters, to future work. 
Specifically, in Table~\ref{tab:ablation}, we report the values of the key evaluation metrics (i.e., cache hit ratio and number of disk reads/writes) for MUSTACHE when $k=\{10, 15, 20, 25, 30, 35, 40\}$.
\begin{table}[ht]
\centering
\begin{tabular}{|c|c|c|c|}
\hline
\textbf{Horizon} ($k$) & \textbf{Hit Ratio} $\uparrow$ & \textbf{\#Reads} $\downarrow$ & \textbf{\#Writes} $\downarrow$\\ 
\hline
$10$ & $0.895$ & $19,854$ & $105,670$ \\ \hline
$15$ & $0.898$ & $19,272$ & $102,017$ \\ \hline
$20$ & $0.900$ & $19,006$ & $99,574$ \\ \hline
$25$ & $0.902$ & $18,746$ & $97,850$ \\ \hline
$30$ & $0.903$ & $18,665$ & $96,516$ \\ \hline
$35$ & ${\bf0.905}$ & $18,602$ & $95,495$ \\ \hline
$40$ & $0.905$ & ${\bf18,435}$ & ${\bf95,156}$\\ \hline
\end{tabular}
\caption{\label{tab:ablation}The impact of the prediction horizon ($k$) on the performance of MUSTACHE. 
}
\end{table}

From this analysis, we may observe that MUSTACHE performs better as the prediction horizon increases until the improvement becomes negligible or even null (i.e., when $k\geq 30$).
This result is compliant with the high predictive accuracy of our page delta forecaster. 
Intuitively, the higher the look-ahead window, the more likely the predictor spots the pages that will be referenced in the future. Thus, removing them from the set of candidates to evict will keep a more accurate set of pages stored in the memory cache. 
\section{Limitations}
\label{sec:limitations}
The original aim of this work was to demonstrate that a page cache replacement algorithm learned from data would be more powerful than existing fixed policies. 
Evidence collected from our experiments has indeed shown that MUSTACHE outperforms traditional baselines.
However, some limitations should be addressed before MUSTACHE can be deployed on real-world OSs.
Amongst those, it is worth mentioning the following.

First, MUSTACHE must perform an offline training step to learn an accurate page (delta) request forecaster. 
Moreover, such a training stage previously requires collecting large sequences of memory accesses, which can be achieved by profiling the system's workload using tools like Intel Pin. 
Data collection and preprocessing, along with model training, can be costly.

Second, model aging might be a severe issue for MUSTACHE: establishing the right frequency for re-training from scratch or even fine-tuning an existing page request forecaster is crucial to balancing cost and performance. 
Indeed, refreshing the model too often would maybe keep high its predictive accuracy at the expense of unsustainable training costs. 
On the other hand, an aged model would be less expensive, but it would degrade the quality of predictions and, thus, the performance of MUSTACHE. 
A typical solution, therefore, is to monitor the cache hit ratio and trigger model re-training as soon as that value falls below a given threshold. 
Anyway, the parameters of the trained model can then
be communicated to the hardware with a new ISA interface~\cite{shi2021asplos}. 

Once deployed, MUSTACHE can query the trained model online at inference time via a lightweight dedicated hardware component for neural network inference. 
For example, Zangeneh et al.~\cite{zangeneh2020micro} use such an approach to improve branch prediction accuracy using CNNs.
\section{Conclusion and Future Work}
\label{sec:conclusion}
In this work, we have presented MUSTACHE, a new page cache replacement policy that uses a multi-step-ahead page forecasting module to reduce the set of pages candidate for eviction. 
The main advantage of MUSTACHE over traditional page cache replacement heuristics (e.g., LRU) is its ability to learn from sequences of memory accesses rather than using a fixed policy. 

We first collected a very large dataset of memory page requests by profiling the execution of a benchmark suite of programs (PARSEC) with a dedicated tool (Intel Pin). 
A subset of this dataset has been used for training different deep learning models for sequential prediction, whose goal was to forecast the next $k$ pages referenced based on the latest $w$ memory accesses observed. 
Specifically, we trained three models: MLP, LSTM, and Encoder-Decoder LSTM. 
We tested offline the predictive accuracy of each model and found that LSTM was the best-performing page request forecaster. 
We used this model as the building block of our proposed MUSTACHE algorithm, and we integrated it into a page cache replacement simulator that already implemented five major heuristics: Random, FIFO, LRU (exact), CLOCK (second chance LRU approximation), and OPT (Bélády's Algorithm). 
In addition, we compare MUSTACHE with another simulator that implements the Adaptive Cache Replacement (ARC) policy.

Experiments demonstrated that MUSTACHE outperformed all the heuristic baselines, improving the cache hit ratio by $1.9\%$ w.r.t. exact LRU and $3.1\%$ w.r.t. CLOCK, halving the gap with the optimal strategy (OPT). 
Moreover, MUSTACHE significantly reduced by $18.4\%$ and $10.3\%$ the number of I/O operations (i.e., reads/writes) required to handle cache misses, respectively.

In future work, we plan to investigate how to make MUSTACHE efficiently deployable on real-world systems. 
Moreover, we will also experiment with more recent, attention-based Transformer architectures for training a more powerful page request forecaster.
Studying the impact of other parameters, such as the number of past memory accesses observed ($w$), on the performance of MUSTACHE is also an interesting direction to explore.
Finally, we may want to consider an alternative formulation of the next-$k$ page request forecasting problem as a ranking task rather than a sequential prediction task, as we presented in this paper.

\bibliographystyle{mlsys2023}

%


\end{document}